\documentclass[%
prl,reprint,twocolumn,
nobibnotes,
 amsmath,amssymb,
 aps,
]{revtex4-1}

\pdfoutput=1 
\usepackage[pdftex]{graphicx}
\usepackage{siunitx}

\usepackage[usenames]{xcolor}
\usepackage{float}
\definecolor{ao(english)}{rgb}{0.0, 0.5, 0.0}
\definecolor{forestgreen(web)}{rgb}{0.13, 0.55, 0.13}

\usepackage[normalem]{ulem}
\usepackage{MnSymbol}
\usepackage{wasysym}

\usepackage{amssymb,amsfonts,amsmath}
\usepackage{calc}

\newcommand{\gdot}{\dot{\gamma}}
\newcommand{\micron}{$\mu$m}
\newcommand{\rs}{s$^{-1}$ }

\begin{document}


\title{A unified description of the rheology of hard-particle suspensions}



\author{B. M. Guy}
\author{M. Hermes}%
\author{W. C. K. Poon}
\affiliation{SUPA, School of Physics and Astronomy, The University of Edinburgh, King's Buildings, Peter Guthrie Tait Road, Edinburgh, EH9 3FD, United Kingdom}
\date{\today}

\begin{abstract} 
The rheology of suspensions of Brownian, or colloidal, particles (diameter~$d \lesssim 1~\mu$m) differs markedly
from that of larger grains ($d \gtrsim 50~\mu$m). Each of these two regimes has been separately studied, but the flow of suspensions with intermediate particle sizes ($1~\mu\textrm{m} \lesssim d \lesssim 50~\mu$m), which occur ubiquitously in applications, remains poorly understood. By measuring the rheology of suspensions of hard spheres with a wide range of sizes, we show experimentally that shear thickening drives  
the transition from colloidal to granular flow across the intermediate size regime. This insight makes possible a unified description of the (non-inertial) rheology of hard spheres over the full size spectrum. Moreover, we are able to  test a new theory of friction-induced shear thickening, showing that our data can be well fitted using expressions derived from it.
\end{abstract}

\maketitle



Complex fluids, polymers, colloids and surfactant solutions find wide applications, partly because of their highly tuneable behavior under deformation and in flow. The success of the mean-field `tube' model for polymers~\cite{Colby2003}, which describes how each chain is constrained by thousands of neighbours, means it has long been possible to predict {\it ab initio} their linear and non-linear rheology from the molecular topology with very few free parameters. In particular, a scaling description is available of the dependence of rheology on molecular weight. 

However, progress in suspension rheology has been more difficult~\cite{Mewis2012}. The small number of nearest neighbours (order 10) rules our any mean-field description: local details matter. It is now possible to predict the low-shear viscosity of a suspension of Brownian hard spheres (HS, diameter $d \lesssim \SI{1}{\micro\meter}$) up to volume fractions of $\phi \lesssim 0.6$, and the rheology of granular HS ($d \gtrsim \SI{50}{\micro\meter}$) is increasingly being studied. Surprisingly, however, how the rheology of HS changes over the whole size spectrum remains unknown, because the behavior in the industrially-ubiquitous intermediate size regime, $1 \lesssim d \lesssim \SI{50}{\micro\meter}$, has not been systematically explored. We offer such an exploration in this Letter, and show that the physics bridging the colloidal and the granular regimes is shear thickening. 



The rheology of colloidal HS is well known~\cite{VanderWerff1989, Krieger1972, Petekidis2004}: the viscosity is determined by the particle volume fraction, $\phi$, and the dimensionless shear rate, or P\'eclet number, Pe~($=\tau_B \gdot$, the shear rate $\dot{\gamma}$ non-dimensionalised by the Brownian time, $\tau_B$, needed for a free particle to diffuse its own radius). At Pe~$\ll 1$ the flow is Newtonian; the viscosity becomes immeasurably large at $\phi_g \approx 0.58$ \cite{Petekidis2004, Phan1996}. Shear thinning starts at Pe~$\lesssim 1$, reaching a second Newtonian regime at Pe~$\gg 1$ with a viscosity that
diverges at random close packing \cite{Mewis2012},
$\phi_{\rm RCP} \approx 0.64$, the densest amorphous packing for lubricated (frictionless) HS. 

Since $\tau_B$ scales as $d^3$, granular HS inhabit the Pe~$\gg 1$
regime at all practical shear rates. Extrapolating na\"ively from the above description of colloidal flow,  one expects Newtonian behaviour with a viscosity diverging at $\phi_{\rm RCP}$. Experiments do find a Newtonian viscosity, but it diverges at a
$\phi$ that is lower than $\phi_{\rm RCP}$ \cite{Boyer2011}, the precise value being dependent on interparticle friction~\cite{Mari2014}. 

How suspension rheology transitions from colloidal to granular behaviour as $d$ increases has not been theoretically predicted, and remains experimentally unclear. Previous experiments on size dependence \cite{VanderWerff1989, Krieger1972, Mewis2000} stayed in the colloidal regime, or
used highly polydisperse systems near the granular limit \cite{Tsai1988}. Thus, no unified description over all particle sizes is yet available. 



\begin{figure*}[t]
\centering
\includegraphics[width=0.80\textwidth]{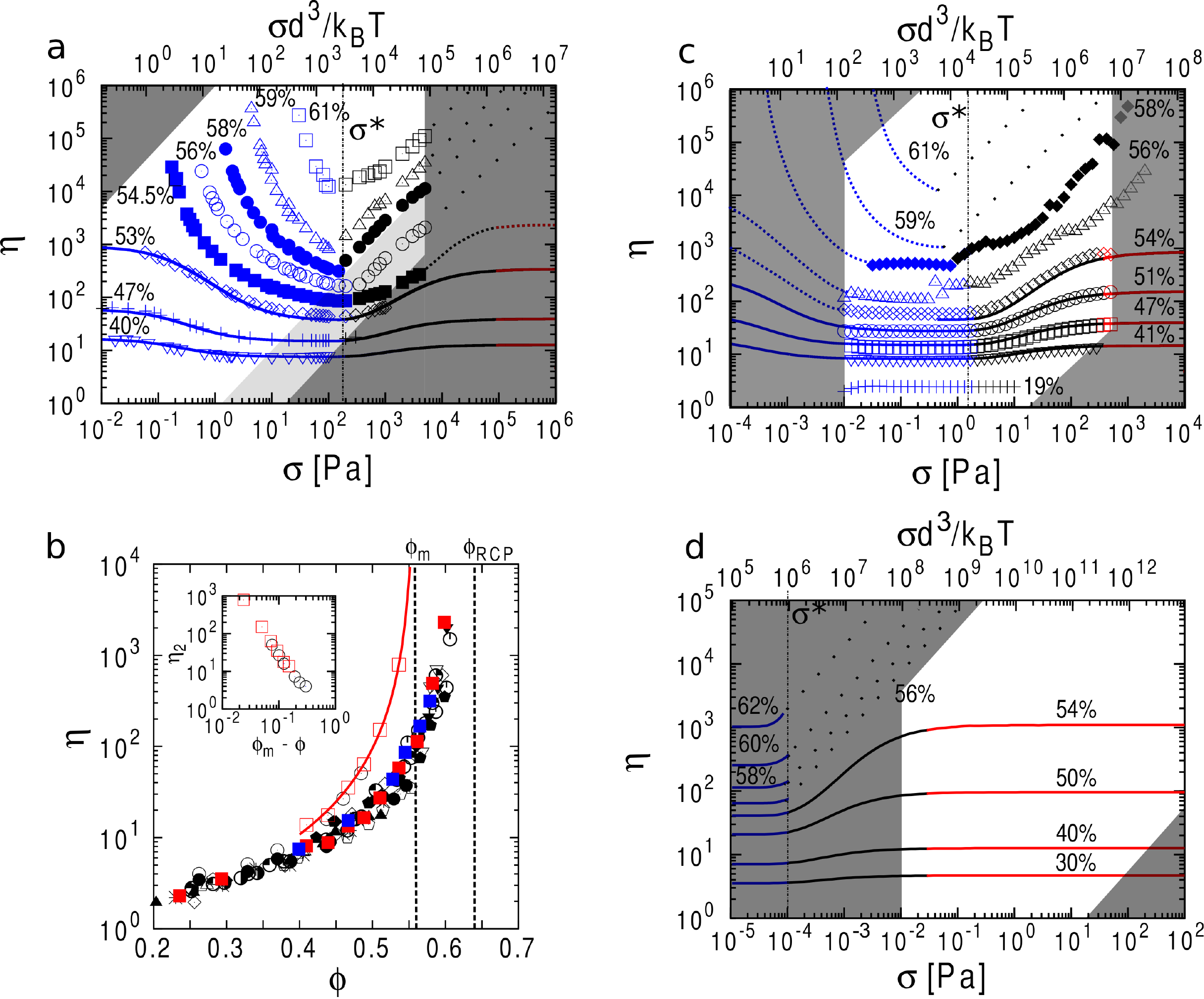}
\caption{Rheology in the colloidal, intermediate and granular size regimes. (a)
Relative viscosity $\eta$ versus shear stress in units of Pa and $k_BT/d^3$ for $d=404$ nm spheres. At $\sigma <10$~Pa we performed downward sweeps in $\gdot$ and at $\sigma >10$~Pa,  upward sweeps in $\sigma$, at volume
fractions (\%)  as labelled. Solid lines: fits based on~\cite{Wyart2014};
finely-dotted lines: schematics based on literature data (with sparsely-dotted = unstable states). Samples shear thicken above a
$\phi$-independent onset stress $\sigma^*$ (vertical dashed line).
Colour scheme: blue = frictionless interactions ($\sigma<\sigma^*$),
black = shear thickening, red = frictional interactions ($\sigma \gg
\sigma^*$). The unshaded region is accessible using
our rheometer, which reaches maximum and minimum shear rates of $8000$~\rs
and $10^{-3}$~\rs\! respectively, and a minimum stress of $10^{-2}$~Pa; the maximum accessible stress is set by a
$d-$dependent fracture stress, $\sigma^\dagger$.
(b) Main:
$\eta(\phi)$ for the the limiting high-shear viscosity,
$\eta_1$, in (a) (\textcolor{blue}{$\blacksquare$}), and the 
lower, $\eta_1(\phi)$ (\textcolor{red}{$\filledmedsquare$}), and upper,
$\eta_2(\phi)$ (\textcolor{red}{$\medsquare$}), branches in (c). Solid red line: least
squares fit to
$\eta_2(\phi) = A(1-\phi/\phi_m)^{-n}$ with
$A=0.20(9)$, $\phi_m = 0.558(5)$ and $n=2.2(2)$. $\eta_1(\phi)$ data 
for other sizes of PMMA spheres in this work: $d = 912$~nm ($\medtriangledown$) and
1800~nm ($\filledmedtriangledown$). Other symbols:
literature high shear viscosities (with $\phi$ shifted by up to $5$\%) for sterically stabilised PMMA
\cite{Phan1996, Petekidis2004, Frith1996, D'Haene1993}, sterically
stabilised silica \cite{VanderWerff1989} and glass beads
\cite{Brown2013}; ($\bullet$) and ($\medcircle$) = lower and upper branches from \cite{D'Haene1993}. Inset:
$\eta_2(\phi)$ versus ($\phi_m - \phi$) including the upper branch from \cite{D'Haene1993}.
(c) $\eta(\sigma)$ for $d  = 3770$~nm spheres, all from upward $\sigma$ sweeps apart from $\phi = 58$\%, which is from a $\sigma$ downward
sweep. The flow in both (a) and (c) was unsteady for $\phi \geq 0.56$, and points
represent temporal averages. 
(d) Schematic for the rheology of $d=350$~\micron\; spheres calculated from theoretical fits to \cite{Wyart2014} in (c).
}
\label{fig:mainfig}
\end{figure*}

\begin{figure}[t]
\centering
\includegraphics[scale=0.50]{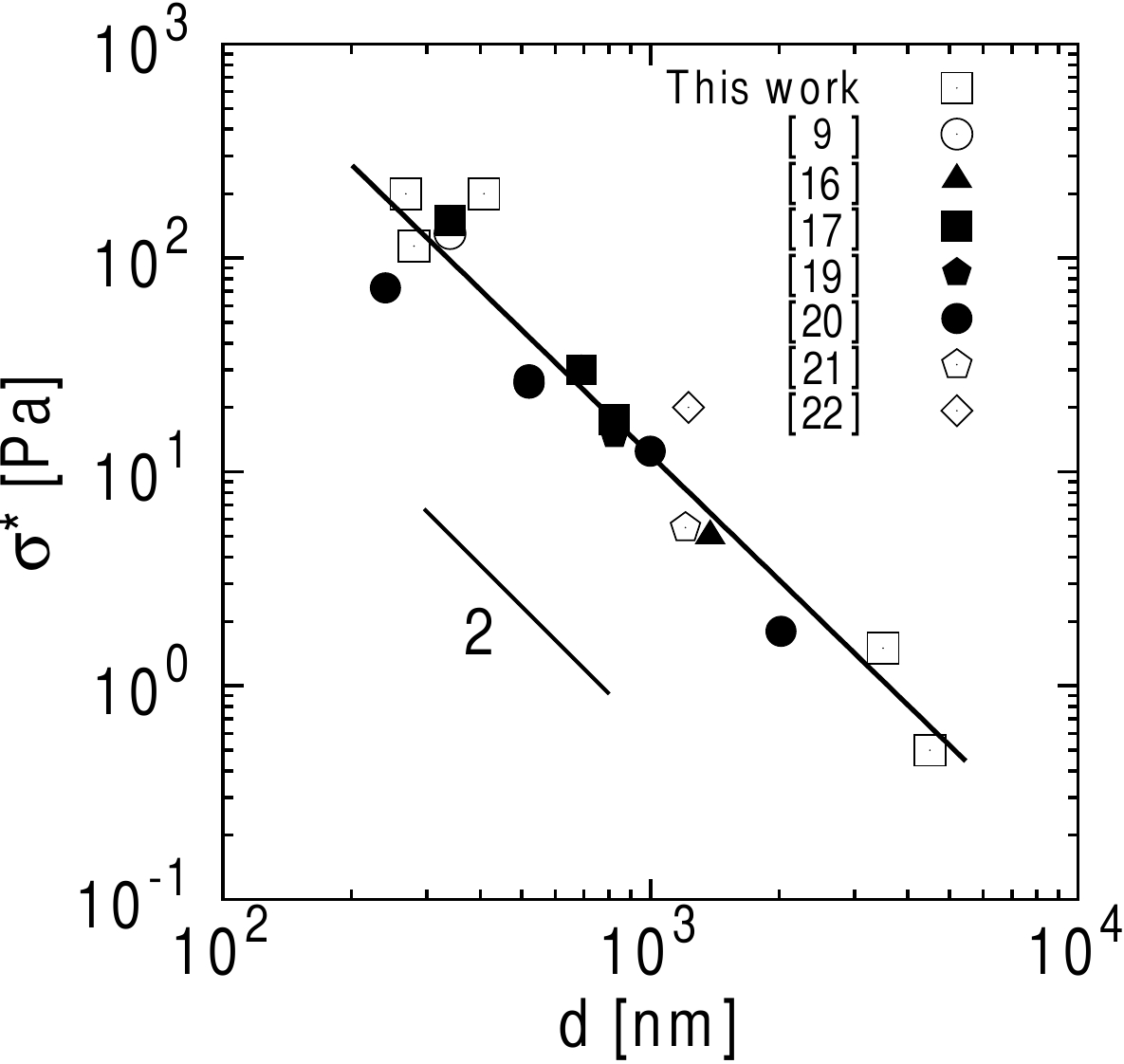}
\vspace{0.5 cm}

\caption{Scaling of the onset stress with particle size. Onset stress
$\sigma^*$ versus particle diameter $d$ for PHSA-stabilised PMMA spheres in this and previous \cite{Mewis2000,  Frith1996, Krishnamurthy2005, Strivens1976, Smith2010,Isa2007} work. The solid line is a least-squares fit to a power law $\sigma^* = Bd^{\alpha}$, with $B=2 \times 10^{-11} \textrm{ J.m}^{-1.1}$
and $\alpha=-1.9(1)$.}
\label{fig:epsilonR}
\end{figure}

\begin{figure}[t]
\centering

\includegraphics[scale=0.28]{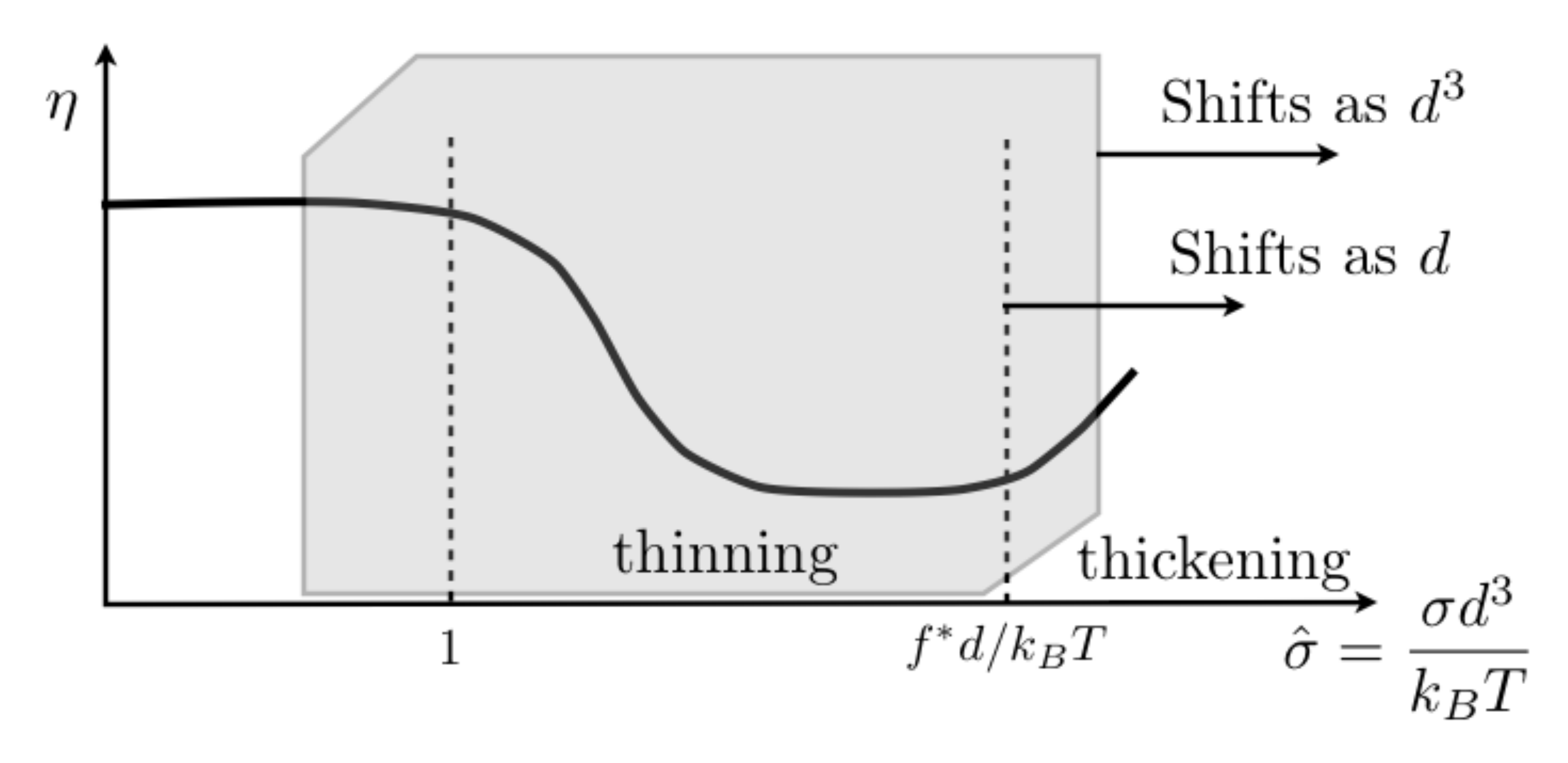}
\vspace*{0.5cm}

\vspace*{-0.5cm}
\caption{Hard-sphere rheology unified. Relative viscosity, $\eta$, versus dimensionless  stress, $\hat \sigma$. The observable window (shaded) for colloids  shows shear thinning, which begins at $\hat\sigma \simeq 1$; shear thickening, beginning at a dimensionless onset stress of $\hat\sigma^* = \sigma^* d^3/k_BT = f^*d/k_BT$, occurs towards the high end of accessible stresses. As the particle diameter $d$ increases, $\hat\sigma^*$ shifts right as $d$, and the second Newtonian regime in the $\eta(\hat\sigma)$ curve 
is stretched out. However, the (shaded) observable window shifts right faster, as $d^3$. Thus, shear thinning becomes unobservable for intermediate particles sizes, and the shear-thickened state fills the observable window of the largest particles (grains). Figure~\ref{fig:mainfig}a, c, d show three snapshots of this scenario.}
\label{fig:schematic}
\end{figure}

We study the rheology of the intermediate size regime using poly-methylmethacrylate
(PMMA) spheres sterically stabilised by
poly-12-hydroxystearic acid (PHSA) with $d = 268$
to 3770~nm, dispersed in a density-matching solvent, Fig.~\ref{fig:mainfig}. We show that the transition from colloidal to granular behaviour is driven by the widespread phenomenon of shear thickening \cite{Brown2013} at a critical `onset stress', whose scaling with particle size ($\propto d^{-2}$), Fig.~\ref{fig:epsilonR}, differs from that of the intrinsic stress scale ($\propto d^{-3}$). Understanding this complexity  leads to a unified description of hard particle rheology over the whole range of $d$, Fig.~\ref{fig:schematic}. Interestingly, our results also confirm recent theory and simulations~\cite{Wyart2014,Mari2014,Seto2013} ascribing shear thickening to the formation of frictional contacts. 

%
%
%

We used polymethylmethacrylate (PMMA) particles stabilised by 5-\SI{10}{\nano\meter}
poly-12-hydroxystearic acid (PHSA) `hairs'~ \cite{Wagstaff1971} in a density-matching mixture of cycloheptylbromide (CHB) or cyclohexylbromide  and decalin. We present data for $d=404$~nm (from
static light scattering) and $3770$ nm (from microscopy) particles with polydispersity $\sim$10\% (from light scattering and electron microscopy, respectively). Data for $d \approx 280$ nm (in decalin), $268$ nm, $912$ nm, $1800$ nm and $4500$ nm (in CHB and decalin) give the same picture. Samples were prepared by diluting a close packed sediment, using simulations~\cite{Farr2009} to estimate $\phi_{\rm RCP}$ of polydisperse HS. The solvent viscosities were $\eta_f=2.83$ mPa.s for
the large and $2.4$ mPa.s for the small particles at 19$^{\circ}$C. Adding an excess of screening salt tetrabutylammonium chloride did not change the rheology;  we present salt-free data.

Rheology was performed in an Anton Paar 301 rheometer in truncated cone and plate geometry (cone angle $1^{\circ}$, radius $25$ mm, truncation gap 100 $\mu$m) in stress-controlled mode unless otherwise stated. A sandblasted cone (surface roughness $\sim \SI{10}{\micro\meter}$) and a base plate roughened with silicon carbide powder (surface roughness $\sim \SI{5}{\micro\meter}$) were used. Even with a solvent trap, artefacts due to drying were evident if samples were left to equilibrate at shear rates $\gdot \lesssim 10^{-2}$ s$^{-1}$. Thus, we worked at $\gdot > 0.01$~s$^{-1}$. 



To establish a baseline, we first explore $d=404$~nm colloids.
Figure~\ref{fig:mainfig}a shows the relative viscosity
$\eta=(\sigma/\gdot)/\eta_f$ (solvent viscosity $\eta_f=2.4$~mPa.s) as a
function of applied shear stress $\sigma$. The region in each $\eta(\sigma)$ plot
in Fig.~\ref{fig:mainfig} {\it not} shaded dark grey is the `observable window'
accessible by conventional rheology, and points are experimental data.
Additionally, the light grey region in Fig.~\ref{fig:mainfig}a is accessible but typically not probed in previous work \cite{VanderWerff1989, Petekidis2004}. 

Our data for $\phi = 0.40$ and $0.47$ show classic shear thinning to a
high-shear Newtonian plateau, $\eta_1(\phi)$, in the observable window. The
$\phi = 0.53$ sample behaves similarly in this window, but shear thickens
beyond it.  At $\phi = 0.56$, we see the onset of discontinuous shear thickening (DST), i.e.~the gradient $d \log{\eta}/\log{\sigma}$ reaches 1, corresponding to a vertical flow curve, $d\gdot/d\sigma=0$. Our $\eta_1(\phi)$ data, Fig.~\ref{fig:mainfig}b
(\textcolor{blue}{$\blacksquare$}), agree with previous measurements of the
`high-shear viscosity' of various colloids \cite{ VanderWerff1989, Petekidis2004, Phan1996, Frith1996, D'Haene1993}.
Within
experimental uncertainty, shear thickening begins at a fixed `onset stress',
$\sigma^* \simeq 200$~Pa, for all $\phi$. At a  higher $\phi$-independent stress
$\sigma^\dagger \simeq 10^5$~Pa, samples
fracture -- this is the high-$\sigma$ boundary to the observable window.
(The lower right, slanted, boundary is due to an inertial instability expelling
samples from the instrument.)

Increasing $d$ by an order of magnitude to $3770$~nm has a
dramatic effect on the rheology in the observable window,
Fig.~\ref{fig:mainfig}c. Now, at $0.41 \leq \phi \leq 0.54$, there is a transition from a lower to a higher, shear-thickened,  Newtonian plateau as $\sigma$
increases \cite{Brown2013,Cwalina2014}. The transition
to a shear-thickened state again begins at a $\phi$-independent onset stress,
now at $\sigma^* \simeq 2$~Pa. Approximately the same onset stress applies at $\phi \geq 0.56$, but now the high-shear plateau vanishes.
The data become noisy and $d\log{\eta}/d\log{\sigma} \simeq 1$, which is the DST
limit. The onset of fracturing in the sample is now at $\sigma^\dagger \simeq
500$~Pa, above which the data show poor reproducibility and strong
history dependence. 


The two plateau viscosities for the $d =3770$~nm particles fall on separate diverging branches,
Fig.~\ref{fig:mainfig}b (\textcolor{red}{$\filledmedsquare$} and
\textcolor{red}{$\medsquare$}) \footnote{Note that below $\phi_m$, sandblasted parallel plate measurements at different gap heights \protect{\cite{yoshimura1988wall}} revealed a small amount of wall slip at all stresses (slip length $\sim 100$ $\mu$m). The corrected viscosity curves $\eta(\sigma)$ are still sigmoidal with an upper (frictional) plateau, and all our conclusions remain unchanged. Imaging revealed massive wall slip in shear thickened samples above $\phi_m$.
}. The lower plateau viscosity here corresponds to $\eta_1(\phi)$ for the $d = 404$~nm colloids (\textcolor{blue}{$\blacksquare$}): both diverge at $\phi \simeq 0.64 = \phi_{\rm RCP}$.
Thus, we infer an experimentally-inaccessible shear thinning regime for these particles at lower $\sigma$, sketched schematically in Fig.~\ref{fig:mainfig}c. The high viscosity branch, $\eta_2(\phi)$, diverges at $\phi_m \simeq 0.55 < \phi_{\rm RCP}$. Interestingly, this is close to random loose packing for frictional granular spheres \cite{Silbert2010}. We infer a corresponding regime of shear-thickening to a plateau viscosity in the $404$~nm particles outside the observable window, Fig.~\ref{fig:mainfig}a. Two viscosity branches can be extracted from previous work on PMMA spheres \cite{D'Haene1993}; each extracted branch can be collapsed onto our corresponding branch by a small shift in $\phi$ (Fig.~\ref{fig:mainfig}b). Note that it would be meaningless (though mathematically possible) to collapse the two branches in Fig.~\ref{fig:mainfig}b into a single curve by a large ($\sim 0.1$) shift in $\phi$: these branches are observed together in the same experiment, and correspond to distinct phenomena.


Recent theory~\cite{Wyart2014} and simulations~\cite{Mari2014, Seto2013}
predict just such a two-branch structure for the viscosity in non-Brownian (granular) suspensions, and
shear thickening is associated with transition from the low to the
high-viscosity branch above an onset stress $\sigma^*$ (see also \cite{Bashkirtseva2009}). At $\sigma \ll
\sigma^*$, particles do not touch and all contacts in the system are
lubricated, while for $\sigma \gg \sigma^*$, all particles are pressed into
frictional contact. The point at which the high-viscosity branch diverges,
$\phi_m$, decreases with increasing (static) friction coefficient, $\mu_p$, between
particles~\cite{Mari2014}, and $\phi_m = \phi_{\rm RCP}$ only if
$\mu_p = 0$. Our data for particles in
the transitional size regime, Fig.~\ref{fig:mainfig}c, are
consistent with this scenario; 
indeed, the solid lines are fits of our data to theory \cite{Wyart2014}
(see Supplemental Material \footnote{See Supplemental Material at [URL]} for details).

The onset stress in this theory, $\sigma^*$, arises physically from the presence of barriers, e.g. from PHSA `hairs' in our case, that stabilise particles against entering the primary van der Waals (vdW) minimum of their mutual interaction potential, and therefore prevent interparticle contact. When the applied stress exceeds a critical value, $\sigma^*$, however, these barriers are overcome and particles are pressed together to make frictional contacts.

We find, Fig.~\ref{fig:epsilonR}, that $\sigma^*$ decreases with particle size as  $\sigma^* \simeq
Bd^\alpha$ with $B = 2 \times 10^{-11}$~J.m$^{-1.1}$ and $\alpha = -1.9(1)$. For charged-stabilised colloids, one expects $\sigma^*\propto d^{-2}$  \cite{Mewis2012}; but charge is irrelevant in our case, because $\sigma^*$ is unchanged by adding salt (data not shown). More relevantly, $\sigma^* \propto d^{-2}$ \cite{Kaldasch2009} and $\propto d^{-1.75}$
\cite{Krishnamurthy2005} scalings are predicted theoretically for stabilising polymer brushes, and 
$d^{-2}$ scaling is found in other sterically-stabilised PMMA particles \cite{Krishnamurthy2005}.

If we take our data as supporting $\sigma^* \propto d^{-2}$, then a constant force $f^*$ is needed to push particles into frictional contact: $\sigma^* \simeq f^*/d^2$ with 
$f^*=3$ $k_BT/\textrm{nm}$. This is comparable to the
$6~k_BT~/\textrm{nm}$ measured between similar PHSA hairs in a different
geometry~\cite{Bryant2002}. The microscopic origins of this constant force $f^* \propto d^0$ are at present unclear (see \cite{Krishnamurthy2005} and Supplemental Material \cite{Note2}), although $\sigma^* \propto d^{-2}$ may be generic: a $d^{-2}$ scaling of a critical onset {\it shear rate} is reported in a review of diverse systems \cite{Barnes1989} (where the substantial data scatter may come from using $\dot\gamma$ rather than $\sigma$ as the scaling variable).


The interplay between the $d^{-2}$ scaling of the onset stress for shear thickening, $\sigma^*$, and the $d^{-3}$ scaling of the intrinsic stress scale controls the colloidal to granular crossover in our system. The progression of observable rheology is set out schematically in Fig.~\ref{fig:schematic} and its caption. (An alternative summary of our findings in the form of a `rheological state diagram' is shown in Supplemental Material Fig.~S1 \cite{Note2}.)
This scenario should be valid for any system in which $\sigma^* \propto d^{-\lambda}$ with $\lambda>0$, with the exact value of $\lambda$ controlling the sharpness of the crossover. Except for $\lambda = 3$, the different scaling of $\sigma^*$ and the intrinsic stress
(always $\propto d^{-3}$ for dimensional reasons) means that a single set of master
curves cannot be found to describe the suspension rheology for all $d$.  


The existence of an `onset stress' $\sigma^{*}$ means that residual vdW
attraction sets a practical limit to the largest particles we could in
principle study. For $d \gtrsim 20$~\micron, the vdW  attraction at the point when the PHSA hairs on neighbouring particles just touch exceeds $\sigma^{*}$ (see Supplemental Material \cite{Note2} for details), and masks the underlying shear thickening \cite{Brown2013}: the system is now a particulate gel. Thus, our system can reach the threshold but cannot properly probe the granular regime of repulsive particle rheology.

While our work is not focussed on shear
thickening {\it per se}, our data impact on the understanding of
this ubiquitous phenomenon. One theory, e.g. \cite{Wagner2009}, ascribes shear thickening to hydrodynamic interactions (HIs) alone, with interparticle friction playing no role. Simulating a system of frictionless spheres with only HIs gives a viscosity increase that is weak and continuous \cite{Brady1997}. It is unclear how the discontinuous shear thickening we observe could arise in this framework, whereas a recent theory of friction-driven shear thickening \cite{Wyart2014} can quantitatively describe our results. 
We also note that the shear thickening we observe is distinct from recent work on inertial
systems \cite{Fernandez2013, Kawasaki2014}, as our particle Reynolds number is at most $
\approx 10^{-2}$, and typically $\lesssim 10^{-4}$ at the onset of thickening.


To conclude, we have shown that the transition from colloidal to granular
rheology is driven by shear thickening. Our data are consistent with recent suggestions \cite{Wyart2014, Mari2014, Seto2013} that shear thickening is associated with the development of frictional
particle contacts at an onset stress, $\sigma^*$, which we find to decrease with
particle size as $d^{-2}$. Thus, particles with $d \lesssim 1~\mu$m will 
behave as frictionless, Brownian hard spheres at most accessible stresses. For intermediate-sized particles, frictionless and frictional
states are observed at low and high stress respectively. Finally, a particulate suspension is granular when $\sigma^*$ is much smaller than commonly-encountered stresses; such a suspension is `always shear thickened'. The size at which this happens depends on the `stabilising force' $f^*$, and therefore the surface chemistry.

That the onset stress is readily accessible and routinely exceedable for a suspension of particles with 1~\micron~$\lesssim d \lesssim$~50~\micron, Figs.~\ref{fig:mainfig}c and \ref{fig:epsilonR}, has significant
practical consequences, especially for concentrations in the range $\phi_m < \phi < \phi_{\rm RCP}$. Once
$\sigma^*$ is exceeded, 
there is no frictional branch with finite viscosity to which the suspension may jump,
Fig.~\ref{fig:mainfig}b. The system shear jams, and shows a qualitative change
in its rheology consistent with previous work on concentrated suspensions
\cite{Lootens2003}. The flow is unsteady, shows edge
fracture and wall slip (confirmed by imaging), and becomes strongly
history-dependent.

Understanding polydisperse industrial suspensions whose particle
size distributions span all three regimes remains a formidable
challenge. Our unified description of particulate rheology over all sizes, summarised in Fig.~\ref{fig:schematic}, has laid the foundation for this challenge to be met. 


\begin{acknowledgments}
This work was funded by the UK EPSRC (EP/J007404/1) and a CASE studentship with Johnson Matthey. We thank Mike Cates, Paul McGuire, Chris Ness, Guilhem Poy and Jin Sun for fruitful discussions, and Andy Schofield for particles.
\end{acknowledgments}
\bibliography{col_gran_paper-2}

\end{document}